\documentstyle{article}
\date{}
\title{On the Influence of Momentum Conservation\\
upon the Scaling Behaviour of Factorial Moments\\
in High Energy Multiparticle Production}
\author{Liu Lianshou, Zhang Yang, Deng Yue\\
{\footnotesize Institute of Particle Physics, Hua-Zhong Normal University,
         Wuhan 430070, China}}

\begin{document}
% \maketitle
\vskip-2cm
\hskip300pt HZPP-96-05

\hskip300pt May 15, 1996
\bigskip

\begin{center}
{\Large On the Influence of Momentum Conservation}
\end{center}
\begin{center}
{\Large upon the Scaling Behaviour of Factorial Moments}
\end{center}
\begin{center}
{\Large in High Energy Multiparticle Production}
\end{center}
\vskip1cm

\begin{center}
{Liu Lianshou, Zhang Yang, Deng Yue\\
{\footnotesize Institute of Particle Physics, Hua-Zhong Normal University,
         Wuhan 430070, China}}
\end{center}
\vskip2cm

\begin{quote}\begin{abstract}
 The experimental results on the falling down of
 scaled factorial moments in azimuthal variable $\phi$
 is studied in some detail. It is shown that this phenomenon
 may be referred to the influence of transverse momentum conservation.
 The existing experimental data from DELPHI, EMU08, NA22 and UA1 are
 successfully explained.
 Various methods are proposed to partly eliminate this influence and
 rule out the 'falling down' of factorial moments.
\end{abstract}\end{quote}

\vfill

\begin{center}
{\bf INSTITUTE OF PARTICLE PHYSICS\\
HUAZHONG NORMAL UNIVERSITY\\
WUHAN \ \ \ CHINA}
\end{center}

\newpage
Since the last decade the anomalous scaling of scaled factorial moments
(SFM's) or intermittency\cite{r1} has been studied extensively\cite{r2} with
the aim of exploring the possible existence of dynamical fluctuation or
multifractal structure of multiparticle spectrum in high energy
collision processes. The corresponding experiments have been performed in
various kind of collisions --- from e$^+$-e$^-$, hadron-hadron, hadron-nucleus
to nucleus-nucleus. Plenty of new and interesting experimental results have
been put in front of us. Various efforts\cite{r3,r4,r5} have been taken to
understand the physics behind these experimental findings.

In this respect, it merits attention that,
in almost all the cases, the logarithm of SFM's versus that of the
cell width $\delta$ rises as the decreasing of $\delta$ (or as the
increasing of partition number $M$ of phase space,
where $M=\Delta / \delta$, $\Delta$ is the total phase
space region in consideration), when $M$ is not very big. There is, however,
a noticeable exception, i.e. the logarithm of SFM's with azimuthal angle $\phi$
as variable fall down as the increasing of ln$M$ in the first few
points\cite{r6,r7,r8,r11}, cf. Fig.1. This prevents us from properly
understanding the physics behind the scaling behaviour of SFM's found
in experiments.

The aim of this paper is to study this phenomenon in some detail.
We will show that the falling down of SFM's in azimuthal angle $\phi$ may
be referred to the influence of momentum conservation (MmCn) constraint in
the collision processes. Various methods will be implemented
to partly eliminate this influence and explore the inherent scaling
property of SFM's.

As is well known\cite{r1}, intermittency refers to the power-law behaviour of
the scaled factorial moments $F_q$ as function of a cell width $\delta$ of
the phase space, in which the hadron multiplicity is measured.
$F_q$ is defined by
\begin{equation}
\label{e1}
 F_q = {1\over M}  \sum _{m=1}^M {\langle n_m(n_m-1)\cdots (n_m-q+1) \rangle
  \over \langle n_m \rangle ^q },
\end{equation}
where $M$ is the number of cells in horizontal average, $n_m$ is the
multiplicity in $m$th $\delta$-cell in a high energy collision event, and
$\langle\ \rangle$ denotes vertically averaging over many events.

However, it should be taken into account that
the total momenta of collision system must keep invariant
in the space-time evolution process of the high energy collisions.
This momentum conservation constraint may create some correlation
between the multiplicity $n_m$ of distinct phase space cells.
But what is the influence of this correlation due to MmCn constraint upon
the values of $F_q$ for different partition number $M$?
In order to attack the solution of this question more directly,
Let us at first take the 1-dimensional (1D) momentum phase space as an example.

Let $\Delta p$ denotes the total momentum district in consideration. The
momentum $p_i$ of each particle is restricted by the conservation
condition $\sum_{i=1}^n p_i=0$,
where $n$ is the total number of particle inside the region $\Delta p$.

Firstly, it is clear that MmCn has no influence on the value
of $F_q$ for $M=1$, because $F_q$ for $M=1$ depends only on the
total number $n$ of particles in the whole region $\Delta p$ no matter how
these particles are distributed, cf. Eq.(~\ref{e1}).
More briefly, $F_q$ is 'blind' to the different distributions of $p$,
therefore it takes the same value regardless of whether there is or isn't MmCn.

When $M=2$ the influence of MmCn comes in force, diminishing the value of
$F_q$. This can be seen as the following. When $M=2$ the total region
$\Delta p$ is divided into two bins with positive and negative $p$
respectively. Let $P_+$ and $P_-$ be the probabilities for a particle to
lie in these two bins respectively. If there were neither dynamical
fluctuation DF (i.e. if the probability did not fluctuate in itself) nor
momentum conservation constraint, these two possibilities would be the same:
 \begin{center}
  $P_+=P_-$ \qquad (in case of neither DF nor MmCn).
  \end{center}
In an individual event, the statistical fluctuation (SF) will make
the multiplicity $n_+$ and $n_-$ of particles with positive and negative $p$
fluctuate,
\begin{center}
 $n_+ \neq n_-$ \qquad (randomly due to SF).
 \end{center}
Assuming SF to be of Bernouli or Poisson type, it can be eliminated by
means of factorial moments method\cite{r1}, resulting in a flat SFM versus $M$, i.e.
\begin{center}
 $F_q^{(M=2)}=F_q^{(M=1)}$ \qquad (in case of neither DF nor MmCn).
 \end{center}
If there is dynamical fluctuation, the $P_+$ and $P_-$ themselves will
be unequal in an individual event, resulting in much larger fluctuations in
$n_+$ and $n_-$, so that $F_q$ will rise from $M=1$ to $M=2$.
On the other hand, The momentum conservation constraint $\sum_{i=1}^np_i=0$
tends to separate particles to the positive and negative $p$ regions,
which reduces the difference between $n_+$ and $n_-$ caused by DF and/or SF,
making $F_q$ smaller than what would be expected without the restriction of
MmCn.

Thus it is predicted that $F_q$ should not rise so steeply as required by DF
or even may fall down from the first point ($M=1$) to the second point ($M=2$)
in contradiction to the flat SFM required by pure SF, because of the
correlation effect between multiplicity $n_+$ and $n_-$ from MmCn.

In Fig.2a,b is shown the results of Monte Carlo simulation of a random
cascading model with and without MmCn. In this model, the total region
$\Delta p$ is divided subsequently with the subdividing ratio $\lambda=2$.
The probability $w_i$ for a subdivision is taken to be
\begin{equation}
\label{e2}
w_{i}={1+\alpha r\over \sum_{i=1}^{\lambda}
           (1+\alpha r_{i})} , \qquad i=1,2,
\end{equation}
where $\alpha$ is the parameter of fluctuation strength, $0\leq\alpha\leq 1$;
$r_{i}$ is a random number in the interval [-1,1]. After $\nu$ steps,
the probability in $k$th subdivided bin ($ k = \{i_1, i_2 \dots, i_\nu\} $)
is
\begin{equation}
\label{e3}
    p(k)=w^{(1)}_{i_{1}}
            w^{(2)}_{i_2} \cdots w^{(\nu)}_{i_{\nu}}.
\end{equation}
The $n$ particles are then assigned to the subdivided bins
according to the Bernoulli distribution
\begin{equation}
\label{e4}
    P(n_1,n_2,\dots,n_K) = {n! \over n_1! n_2! \cdots n_K!}
      p(1)^{n_1}p(2)^{n_2}\cdots p(K)^{n_K},
\end{equation}
where $K=2^\nu$ is the total number of bins. Taking randomly the momenta of
the $n_k$ particles in the $k$th bin ($k=1,2,\dots,K$), a Monte Carlo sample
of particle momenta is obtained. In the calculation we have taken the model
parameter $\alpha=0.5$, the cascading has been performed up to $\nu=7$th step.

In this model the dynamical fluctuation is taken into account up to the
$\nu$th step of random cascading, i.e. up to the bin width as small as
$\delta p = \Delta p/2^\nu$. When $\nu$ is large, $\delta p$ is very small,
the neglecting of further DF will cause trivial error in the final results.

>From the construction of model the resulting SFM's have the property of
anomalous scaling
\begin{equation}
\label{e5}
   F_q(M) \propto M^\varphi,
\end{equation}
so that the log-log plot of ln$F_q$ versus ln$M$ is a straight line
pointing upward for the point of $M=2^\nu$\footnote{The 'chain effect'
may appear in the plot of $\ln F_q$ versus $\ln M$ if the continuously
diminishing scale is taken in this kind of random cascading
model\cite{liu1}.}, cf. Fig.2$a$.

In the above model there is, of course, not any restriction from
MmCn in the resulting sample, i.e. $\sum_{i=1}^n  p_i\neq 0$. We use a
simple method to put in the constraint of MmCn. Let
\begin{equation}
\label{e6}
       p_c={1\over n}\sum_{i=1}^n  p_i,
\end{equation}
and shift the momentum of every particle in the event by an amount $p_c$
\begin{equation}
\label{e7}
  p'_i=p_i-p_c.
\end{equation}
Then the new momenta $p'_i$ obey the conservation law, $\sum_{i=1}^n p'_i=0$.
This transform does effect the distribution of $p$, cf. Fig.3,
but not very much, and at the same time the fractal structure
existing in the original sample retains in the new one.
Using the cumulative variable as usual\cite{r9},
the distribution becomes flat again and the SFM's can be calculated.

The results are shown in Fig.2$b$. It can be seen in the figure that the
first point, corresponding to $M=1$ has the same height as that of Fig.2$a$
while the second point, $M=2$, is lower than that of Fig.2$a$ due to MmCn,
resulting in a falling down of ln$F_2$ from the first to the second point,
confirming the prediction given above. When $M$ increases further,
ln$F_2$ increases again due to dynamical fluctuation.

Thus we see that the falling down of $F_q$ at the first few points is due
to putting in force of the MmCn effect. If we take the positive and
negative $p$ particles as particles of two "events", calculating SFM's in
these "events" separately and then take the vertical average over them,
the falling down phenomenon will disappear. It is really the case in our
model, cf. Fig.2$c$.

On the other hand, if we make a cut and consider only the particles with
$|p|<p^{\rm cut}$, then the remaining particles will no longer be restricted
so strongly by MmCn and the falling down phenomenon may graduately disappear.
In Fig.2$d$ is shown the result of our model with $p^{\rm cut}=0.8$,
a straight line is recovered.

Let us now turn to the more realistic case of 3-dimensional (3D)
phase space. It is noticeable that the momentum conservation constraint
of sample is very different in transverse and longitudinal phase space.
Generally, not all final fragments of high energy collisions are considered
in the intermittency experiments. In particular, the leading particles,
which take away in the average $50\%$ of longitudinal momenta, are neglected.
Therefore, the restriction of
longitudinal momentum conservation on the particles analysed is quite weak.
In fact, the experimental data of SFM's in longitudinal phase space (e.g.
rapidity $y$ or pseudorapidity $\eta$) do not fall down in the plot of
$\ln F_q\sim\ln M(y)$, cf. Ref.\cite{r2}.

On the other hand, the constraint effect of MmCn in transverse plan is a
little different between the samples from the different experiments.
It depends on the distinct experimental conditions,
i.e. the analysed phase space district cover both the central and fragmental
regions, or only the central region, or just part of the central region.

There is usually no cut in transverse momentum in the fixed target experiments
and the e$^+$e$^-$ collisions, and the transverse momenta of leading
particles are almost zero. So if the produced particles in both central
and fragmental region are considered (e.g. DELPHI collaboration\cite{r8}),
the transverse momentum conservation must be satisfied very well.

In some intermittency experiments, only final state hadrons of central
region are analysed. For example, the NA22 collaboration put a cut of
rapidity $|y|\le 2$ in longitudinal direction\cite{r6} although there
is no cut in the transverse directions. So the transverse momenta of
final hardons in fragmental region have been neglected.
However, since many experimental
finding had shown that the final particles in central and fragmental
regions may come respectively from several relatively independent
sources (or fire balls)\cite{liu}, the transverse MmCn condition
is also kept approximately for the particles of central region alone.

In UA1 experimental data\cite{r11}, only a part of the particles in the
central region are analysed. A pseudorapidity cut of $|\eta |\le 1.5$ is taken
in addition to the beam pipe cut of $p_t>0.15$GeV in
transverse direction. So the influence of transverse MmCn in UA1
experimental data is not so strong, which will manifest itself by
the weaker falling degree of SFM's in the plot of $\ln F_q\sim\ln M(\phi )$.

In order to reveal the influence of transverse MmCn upon the anomalous
scaling behaviour of SFM's in 3D phase space, we generalize the
above 1D cascading model with MmCn into 3D phase space.
Constructing a 3D random cascading model in the space of
$0\le y\le 1,\ 0\le p_t\le 1,\ 0\le\phi\le 2\pi$, the results of Monte Carlo
simulation are shown in the first column of Fig.5.
The 3D ($y,p_t,\phi$) curve is a straight line as required by the cascading
model, and the 1-, 2-dimensional SFM's get saturated,
which is referred to the well known projection effect\cite{r10}.

Now let us impose the transverse MmCn condition
on the random cascading model. After all the vector momentum $\vec p_i$
of the produced particles are determined in a simulated event, we shift
the vector transverse momentum $\vec p_{t_i}$ of every particle by
an amount $\vec p_{t_c}={1\over n}\sum_{i=1}^n\vec p_{t_i}$, i.e.
\begin{equation}
\vec {p'}_{t_i}=\vec p_{t_i}-{1\over n}\sum\limits_{j=1}^n\vec p_{t_j}.
\end{equation}
Then the transverse MmCn condition
\begin{equation}
\sum\limits_{i=1}^n\vec{p'}_{t_i}=0
\end{equation}
is satisfied.

The distribution function of rapidity $y$ is, of course, not influenced
by this transform (not shown in figure). The effect of the translation
upon the distribution of
$p_t$ and $\phi$ are shown in Fig.4. It can be seen from the figure
that the $\phi$ distribution is almost uninfluenced by the transform
due to the axial symmetry. So the cumulative variable has to be used
only for $p_t$.

The results of calculation with MmCn put in in this way are shown in the
second column of Fig.5. It can be seen clearly that the falling down
effect does take place for variable $\phi$.
It is also interesting to notice that the minimum of SFM in the
ln$F_2$ versus ln$M$ plot occurs at $M(\phi)=4$,
contradicting the 1D MC results with minimum
location at $M(\phi)=2$. This is as expected, because $M(\phi)=4$ means
that the positive and negative transverse momenta in both
directions (denoted by $p_{ty}$ and $p_{tz}$ respectively) are divided
into different bins, while in the case of smaller $M(\phi)$, $M(\phi)=2$
for example, the positive and negative
transverse momentum in only one direction (say $p_{ty}$) are separated.
Therefore,
in the case of $M(\phi)=2$, $F_q$ is 'blind' to the different distributions
in $p_{tz}$ and the effect of reduction of DF by transverse MmCn is
weaker correspondingly. This is why $F_q(M=1)>F_q(M=2)>F_q(M=4)$.

If we consider the four quadrants as four different "events" and then
take the vertical average, the falling down phenomenon will disappear,
as can be seen in the
result of model calculation shown in the third column of Fig.5.

If we impose a cut on $p_t$ to weaken the influence of MmCn, and consider
only particles with $p_t<p_t^{\rm cut}$, then the falling
down effect gets weakened too and disappear at about $p_t^{\rm cut}=0.7$,
cf. Fig.6.

Thus we have successfully explained the experimental finding on the falling
down of the first few points in the $\ln F_q(\phi)$ versus $\ln M(\phi)$ plot,
cf. Fig.1. The transverse momentum conservation is satisfied in case that all
the produced particles in both central and fragmental regions\cite{r7,r8} are
considered, or approximately satisfied if only those in central region are
considered. According to the above
argument the minimum of $\ln F_q$ versus $\ln M(\phi )$ should occur at
about $M(\phi)=4$. It does occurs at about $M(\phi)=5$ in the experimental
data in Fig.1a,b. On the contrary, in the $p\bar p$ collider experiments
of UA1 collaboration, only a part of the produced particles in central region
are analysed, and there is a transverse $|p_t|>0.15$GeV cut due to the
existence of beam pipe. So the effect of transverse MmCn is much weaker,
cf. Fig.1c and Fig.6c.

Although there is a difference between these two cases,
i.e. in our model of Fig.6c we have imposed a cut at
the upper side of $p_t$, while in UA1 experiment (Fig.1c),
the cut is imposed at the lower side, the physics is the same,
both are the relaxation of the restriction of transverse MmCn due to the
particles carrying only a part of transverse momenta. It is noticeable
that, besides the weakening of the falling down of SFM, this effect also
makes its minimum move to a smaller value of $M$, locating at $M<4$.
This is the case both for the Monte Carlo model and for the experimental
data, cf. Fig.1c and Fig.6c.

We conclude that in order to explore the physics behind the experimental
phenomena on the scaling behaviour of scaled factorial moments (intermittency
phenomena), the influence of transverse momentum conservation must be taken
into account. From the constraint of transverse momentum conservation,
the falling-down of SFM with azimuthal angle variable in DELPHI, EMU08,
NA22 and UA1 can be explained successfully.

Using the method of "quadrant analysis", i.e. taking the
particles in different quadrant as different "event", or imposing a $p_t$
cut, this influence can be partly eliminated, especially, the falling down
of the first few points can be ruled out. So these methods are helpful
to reveal the inherent scaling behaviours of a fractal system.

\bigskip
\noindent{\Large \bf{Acknowledgement} \rm}
\bigskip

One of authors (Z.Y.) wishes to thank Dr. Liu Feng for helpful discussions
about the UA1 experimental sample.
This work is supported in part by the NNSF of China.

%\newpage
\vfill.5cm

\newpage
\noindent{\bf \Large Figure Captions}
\bigskip
\begin{description}
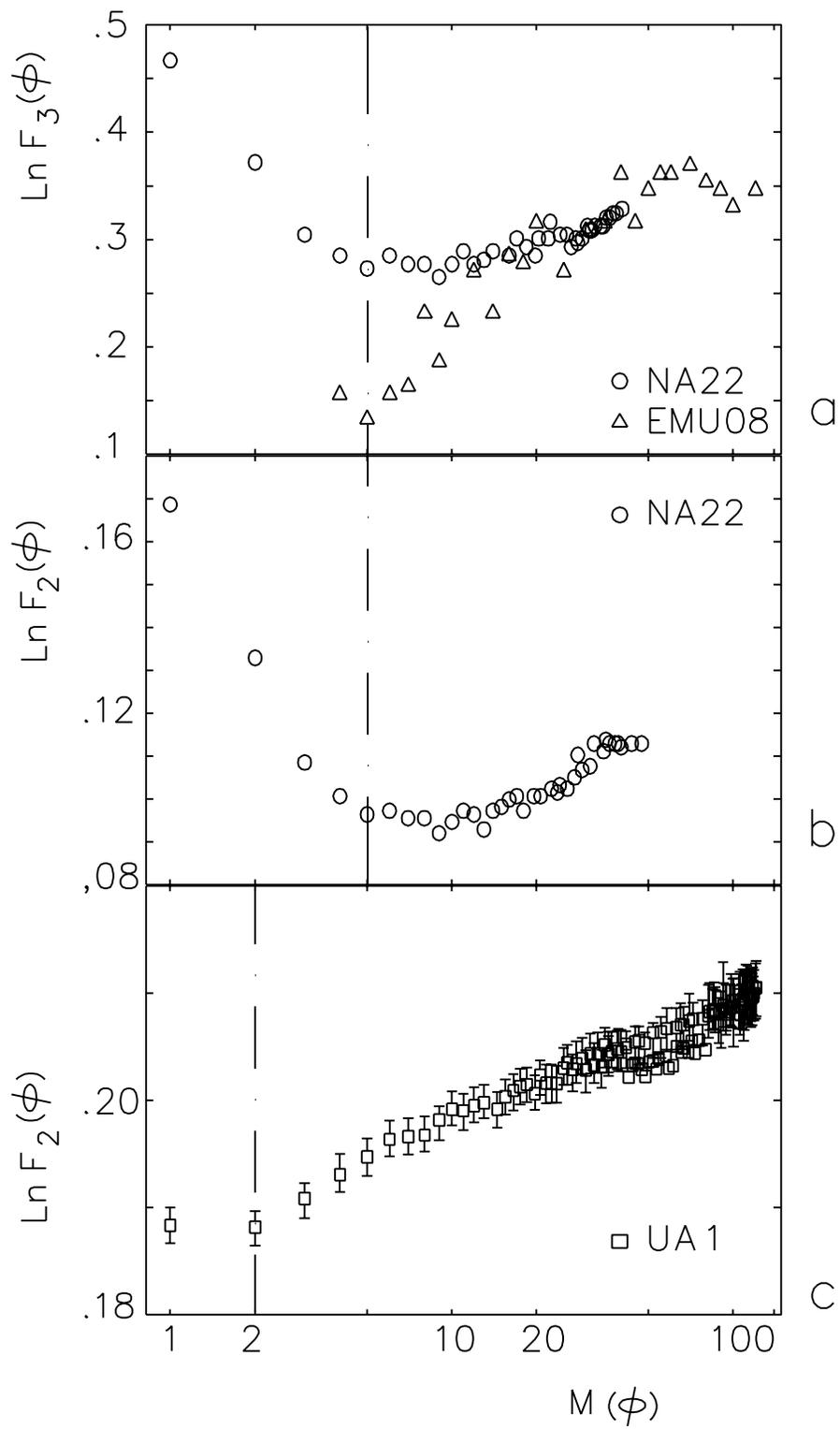
\item[Fig.1]
Scaled factorial moments with azimuthal angle as variable, showing the
'falling down' phenomenon. Data taken from Ref's.\cite{r6,r7,r11}.
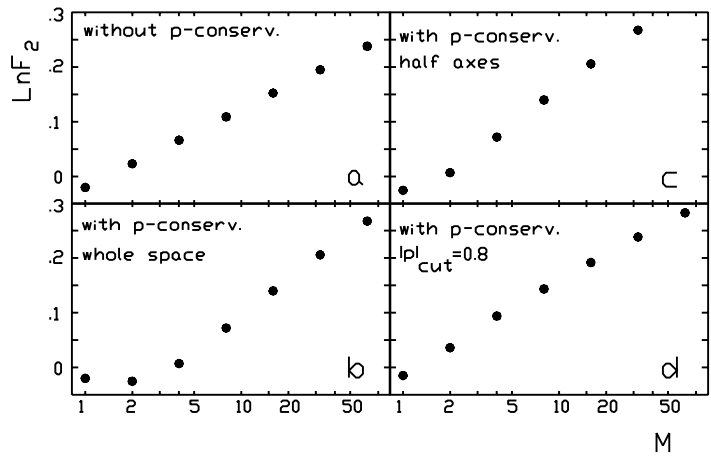
\item[Fig.2]
 The scaling behaviours of SFM's in one-dimensional random cascading model:\\
$\begin{array}{l}
\hbox{\rm (a), original cascading model without MmCn constraint;}\\
\hbox{\rm (b), the cascading model with MmCn considered;}\\
\hbox{\rm (c), the result of 'quadrant analysis' to the cascading model with
               MmCn;}\\
\hbox{\rm (d), the result of the cascading model with MmCn after a cut of}
\ |p|\le 0.8 \ \hbox{\rm is imposed.} \end{array}$
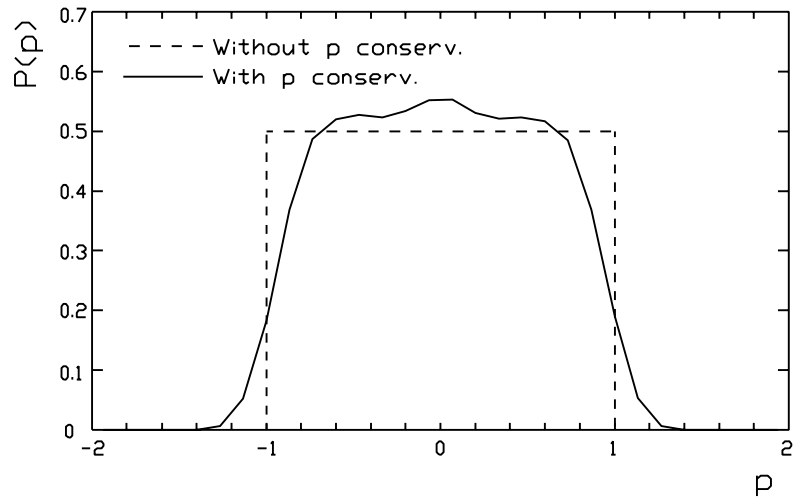
\item[Fig.3]
Distribution of $p$ in 1D cascading model with and without MmCn.
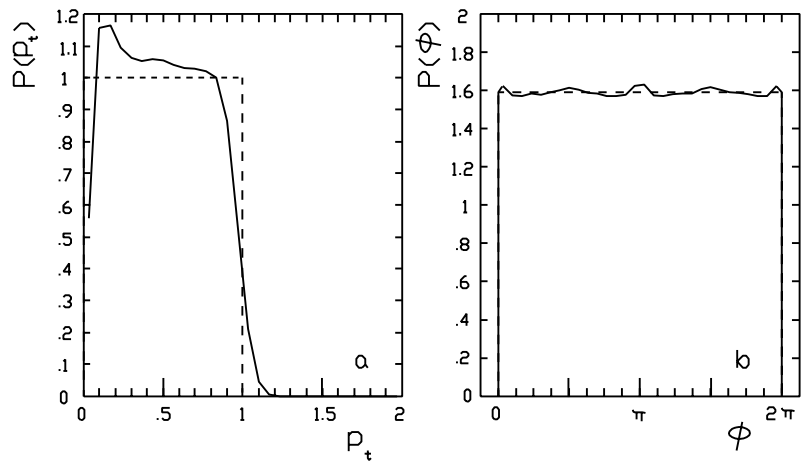
\item[Fig.4]
The influence of shift of transverse momenta upon the distribution function
of $p_t$ and $\phi$ in 3D ($y, p_t, \phi$) phase space. The solid curve
denote the distribution function after MmCn is inputted, and the dashed ones
denote that without MmCn.
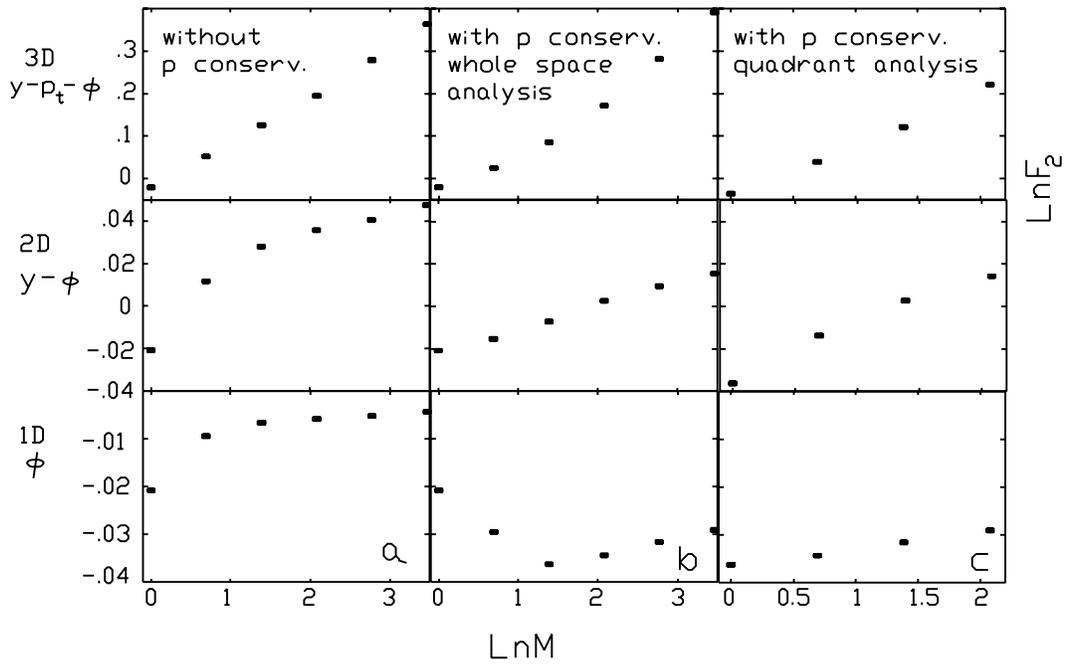
\item[Fig.5]
The scaling behaviours of SFM's and their lower dimensional projection
in 3D cascading model, where the first row is results of 3D ($y-p_t-\phi$)
phase space, the second and third rows are that of 2D ($y-\phi$) and
1D ($\phi$) projections respectively\\
$\begin{array}{l}
\hbox{\rm (a), the original cascading model without MmCn (the first column);}\\
\hbox{\rm (b), the cascading model with MmCn (the second column);}\\
\hbox{\rm (c), the result of 'quadrant analysis' for the cascading model with
MmCn}\\
\hbox{\rm \ \ \ (the third column).}
\end{array}$
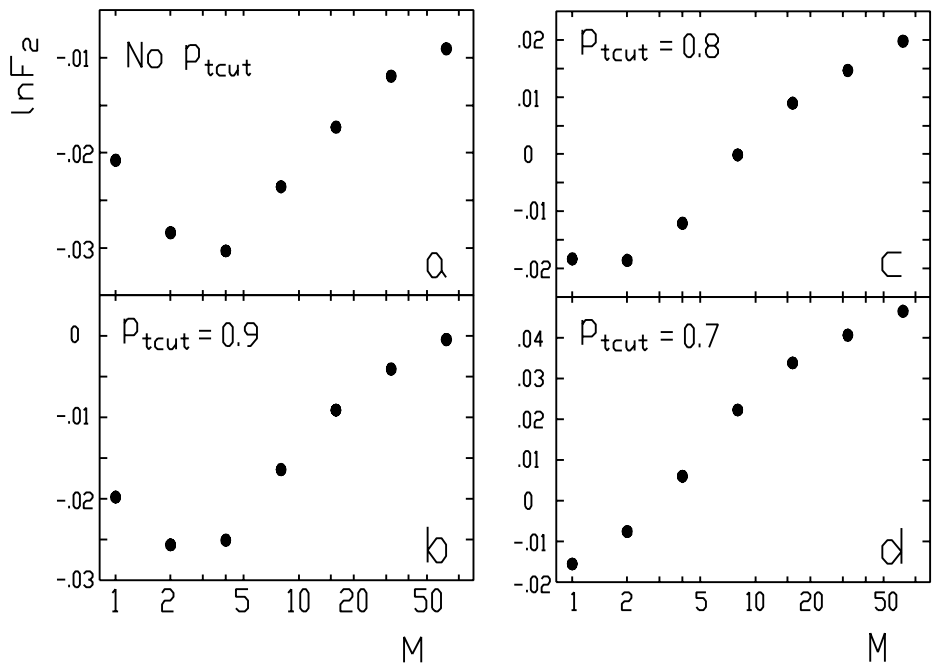
\item[Fig.6]
The results of the 3D cascading model with MmCn after a cut of
$|p|\le p_t^{\rm cut}$ is imposed, (a)-(d) correspond to different
$p_t^{\rm cut}$ values respectively.

\end{description}
\end{document}